\documentclass[sigconf]{acmart}
\AtBeginDocument{%
  }

\setcopyright{acmlicensed}
\copyrightyear{2018}
\acmYear{2018}
\acmDOI{XXXXXXX.XXXXXXX}
\acmConference[Conference acronym 'XX]{Make sure to enter the correct
  conference title from your rights confirmation email}{June 03--05,
  2018}{Woodstock, NY}
\acmISBN{978-1-4503-XXXX-X/2018/06}

\renewcommand\footnotetextcopyrightpermission[1]{}
\usepackage{booktabs}
\usepackage{multirow}
\usepackage{algorithm}
\usepackage{algorithmic}
\usepackage{enumitem}
\usepackage{xcolor}
\usepackage{subfigure}
\usepackage{tabularx}

\usepackage{makecell}
\usepackage{siunitx}    % 可选，对数字进行对齐
\usepackage{graphicx}   % 若需要 \resizebox

\begin{document}

\title[RankMixer: Scaling Up Ranking Models in Industrial Recommenders]{RankMixer: Scaling Up Ranking Models in  \\Industrial Recommenders}

\author{Jie Zhu*, Zhifang Fan*, Xiaoxie Zhu*, Yuchen Jiang*, Hangyu Wang, Xintian Han, Haoran Ding, Xinmin Wang, Wenlin Zhao, Zhen Gong, Huizhi Yang, Zheng Chai, Zhe Chen, Yuchao Zheng, \\ Qiwei Chen$^{\dagger}$, Feng Zhang, Xun Zhou, Peng Xu, Xiao Yang, Di Wu, Zuotao Liu}

\affiliation{
	\country{ByteDance\\
    \{zhujie.zj, zhuxiaoxie.777, jiangyuchen.jyc, wanghangyu.123, hanxintian, dinghaoran, xinmin.wang,
zhaowenlin, gongzhen.666, yanghuizhi, chaizheng.cz, chenzhe.john, zhengyuchao.yc, feng.zhang, zhouxun, xupeng, wuqi.shaw, di.wu, michael.liu\}@bytedance.com, \{fanzhifangfzf, chenqiwei05\}@gmail.com}}

\thanks{* Equal Contributions. \\ $\dagger$ Corresponding authors.}

\renewcommand{\shortauthors}{Zhifang Fan et al.}

\begin{abstract}
  Recent progress on large language models (LLMs) has spurred interest in scaling up recommendation systems, yet two practical obstacles remain. First, training and serving cost on industrial Recommenders must respect strict latency bounds and high QPS demands. Second, most human-designed feature-crossing modules in ranking models were inherited from the CPU era and fail to exploit modern GPUs, resulting in low Model Flops Utilization (MFU) and poor scalability. We introduce RankMixer, a hardware-aware model design tailored towards a unified and scalable feature-interaction architecture. RankMixer retains the transformer’s high parallelism while replacing quadratic self-attention with multi-head token mixing module for higher efficiency. Besides, RankMixer maintains both the modeling for distinct feature subspaces and cross-feature-space interactions with Per-token FFNs. We further extend it to one billion parameters with a Sparse-MoE variant for higher ROI. A dynamic routing strategy is adapted to address the inadequacy and imbalance of experts training. Experiments show RankMixer's superior scaling abilities on a trillion-scale production dataset. By replacing previously diverse handcrafted low-MFU modules with RankMixer, we boost the model MFU from \textbf{4.5\%} to \textbf{45\%}, and scale our online ranking model parameters by two orders of magnitude while maintaining roughly the same inference latency. We verify RankMixer's universality with online A/B tests across two core application scenarios (Recommendation and Advertisement). Finally, we launch 1B Dense-Parameters RankMixer for full traffic serving without increasing the serving cost, which improves user active days by \textbf{0.3\%} and total in-app usage duration by \textbf{1.08\%}.

\end{abstract}







\maketitle

\section{Introduction}

Recommender System (RS) is essential in the process of distributing information. As a significant machine learning scenario, RS predicts users' behaviors towards items based on a large amount of multi-field feature data, including numerical features such as diverse statistics, categorical features such as user and item IDs, user behavior features and content features~\cite{zhang2021deep,lin2023can}. The state-of-the-art recommendation methods are based on Deep Learning Recommendation Models (DLRMs), which flexibly capture feature interactions based on neural networks as the dense interaction layers above the input embeddings from features. The Dense interaction layer in DLRM is critical for RS performance and diverse model structures are proposed~\cite{covington2016deep,din,guo2017deepfm,wang2021dcn,zhangwukong}.

Driven by advancements in Large Language Models (LLMs) that benefit from increasing parameters~\cite{achiam2023gpt,kaplan2020scaling,hoffmann2022training}, scaling up DLRMs to take full advantage of the volume of data is an urgent need. 
Previous research has yielded numerous outcomes on the scaling DLRMs, early studies~\cite{zhang2024scaling,ardalani2022understanding,chitlangia2023scaling} just widen or stack feature interaction layers without modifying the structure. The benefits achieved in this way are modest and occasionally negative~\cite{wang2017deep,lian2018xdeepfm}. Then the follow-up efforts, such as DHEN~\cite{zhang2022dhen} and Wukong~\cite{zhangwukong}, focus on designing innovative DNN structures to boost the scaling performance. However, leveraging model scale to improve performance in recommendation presents unique practical challenges. Unlike NLP or vision tasks, industrial-scale recommendation systems must strictly adhere to tight latency constraints and support extremely high QPS (queries per second). As a result, the core challenge lies in finding a sweet spot that balances model effectiveness and computational efficiency.

Historically, the architecture of ranking models in recommendation has been shaped by CPU-era design principles. These models typically relied on combining heterogeneous diverse handcrafted cross-feature modules to extract feature interactions, but many of their core operators are memory-bound rather than compute-bound on modern GPUs, leading to poor GPU parallelism and extremely low MFU(Model Flops Utilization) often in the single-digit percentage range.Moreover, since the computational cost of CPU-era models was approximately proportional to the number of parameters, the potential ROI from aggressive scaling—as suggested by scaling laws—was difficult to realize in practice.

In summary, the research on scaling laws of DLRMs must overcome the following problems:
\begin{itemize}[leftmargin=10pt]
    \item Architectures should be hardware-aligned, maximizing MFU and compute throughput on modern GPUs.

    \item The model design must leverage the characteristics of recommendation data, such as the heterogeneous feature spaces and personalized cross-feature interactions among hundreds of fields.
\end{itemize}

To address these challenges, we propose a hardware-aware model design method, RankMixer. The core design of RankMixer is based on two scalable components: 1. \textit{Multi-head token mixing} achieves cross-token feature interactions only through the parameter-free operator. This strategy outperforms the self-attention mechanism in terms of performance and computational efficiency. 2. \textit{Per-token feed-forward networks (FFNs)} expand model capacity significantly and tackle inter-feature-space domination problem by allocating independent parameters for different feature subspace modeling. These FFNs also align well with the recommendation data patterns, enabling a better scaling behavior.
To further boost the ROI of large-scale models, we extend the per-token FFN modules into a \textit{Sparse Mixture-of-Experts (MoE)} structure. By dynamically activating only a subset of experts per token for different data, we can significantly increase the model capacity with a minimal increase in computational cost. 
RankMixer adopts a highly parallel architecture similar to transformers, but overcomes several critical limitations of self-attention based feature-interaction: low training efficiency, the combinatorial explosion when modeling cross-space ID similarity and severe memory-bound caused by attention weights matrix. At the same time, RankMixer offers greater model capacity and learning capability under the same Flops compared with Vanilla Transformer.

In production deployment on Douyin’s recommendation system, we demonstrate that it is feasible to scale model parameters by over 100× while maintaining shorter inference latency compared with previous baseline. This is made possible by the RankMixer architecture’s ability to decouple parameter growth from FLOPs, and decouple FLOPs growth from actual cost through high MFU and engineering optimization.

The main contributions can be summarized as follows:
\begin{itemize}[leftmargin=6pt]
    \item We propose a novel architecture called RankMixer follows a hardware-aware model-design philosophy. We design the multi-head token mixing and per-token FFN strategies to capture heterogeneous feature interactions efficiently, and using dynamic routing strategy to improve the scalability of SparseMoE in RankMixer.
    \item Leveraging levers of high MFU and performance-optimization, we scale the model parameters by 70× without increasing the inference cost, including improving MFU and Quantization.
    \item We conduct extensive offline and online experiments and investigate the model's scaling law on the trillion-scale industrial recommendation dataset. The RankMixer model has been successfully deployed on the Douyin Feed Recommendation Ranking for full-traffic serving, achieving 0.3\% and  1.08\% increase on active days and App duration, respectively.
\end{itemize}

\section{Related Work}
\label{sec:related work}

Modern recommendation systems are based on Deep Learning Recommendation Models (DLRMs) and how to effectively model the feature interactions is a crucial factor for DLRMs~\cite{cheng2016wide,guo2017deepfm,huang2022neural,qu2018product,wang2024flip,wang2021dcn,zhang2022dhen}. Wide\&Deep~\cite{cheng2016wide} is one of the earliest efforts, which combines the logistic regression (wide part) and DNN (deep part) to capture low-order and high-order feature interactions respectively. DeepFM~\cite{guo2017deepfm} is another achievement, which integrates the Factorization Machine (FM) and the DNN. In addition, DeepCross~\cite{shan2016deep} is an extension of the residual network~\cite{he2016deep}, aiming to learn automatic feature interactions implicitly. While it has proven to be quite challenging to simply rely on DNNs to learn high-order feature interactions~\cite{qu2018product,rendle2020neural}.
Explicit cross methods design different operators to explicitly capture high-order feature interactions, such as PNN~\cite{qu2018product}, DCN~\cite{wang2017deep} and its successor DCNv2~\cite{wang2021dcn}, xDeepFM~\cite{lian2018xdeepfm}, FGCNN~\cite{liu2019feature} and FiGNN~\cite{li2019fi}. AutoInt~\cite{song2019autoint} and Hiformer~\cite{gui2023hiformer} employ attention mechanism with residual connections to learn complex interactions. DHEN~\cite{zhang2022dhen} proposes to combine multiple interaction operators together. Despite improving accuracy, these new architectures increase the model latency and memory consumption and their model size is relatively small.

Scaling law has emerged as a fundamental theme in deep learning and a pivotal catalyst for numerous breakthroughs over the past decade, particularly in Natural Language Processing (NLP)~\cite{kaplan2020scaling,hoffmann2022training}, Computer Vision (CV)~\cite{dosovitskiy2020image,zhai2022scaling}, and multi-modality modeling~\cite{radford2021learning,ramesh2021zero}, which describe the power-law correlations between model performance and scaling factors, such as model size, data volume, and computational capacity~\cite{kaplan2020scaling,achiam2023gpt,touvron2023llama,team2024gemma}.
Recently, scaling laws in recommendation systems have drawn much attention from researchers~\cite{guo2024scaling}. Studies have explored the scaling strategy for pre-training user activity sequences~\cite{chitlangia2023scaling}, general-purpose user representations~\cite{shin2023scaling,zhang2024scaling} and online retrieval~\cite{fang2024scaling,wang2024scaling}. Wukong~\cite{zhangwukong} stacks FMs and LCB to learn the feature interactions. Orthogonally, \citet{zhang2024scaling} scaled up a sequential recommendation model to 0.8B parameters. HSTU~\cite{zhaiactions} enhances the scaling effect of Generative Recommenders (GRs) which focus more on the sequence part.

\section{Methodology}
\label{sec:methods}

\begin{figure*}
    \centering
    \includegraphics[width=0.85\linewidth]{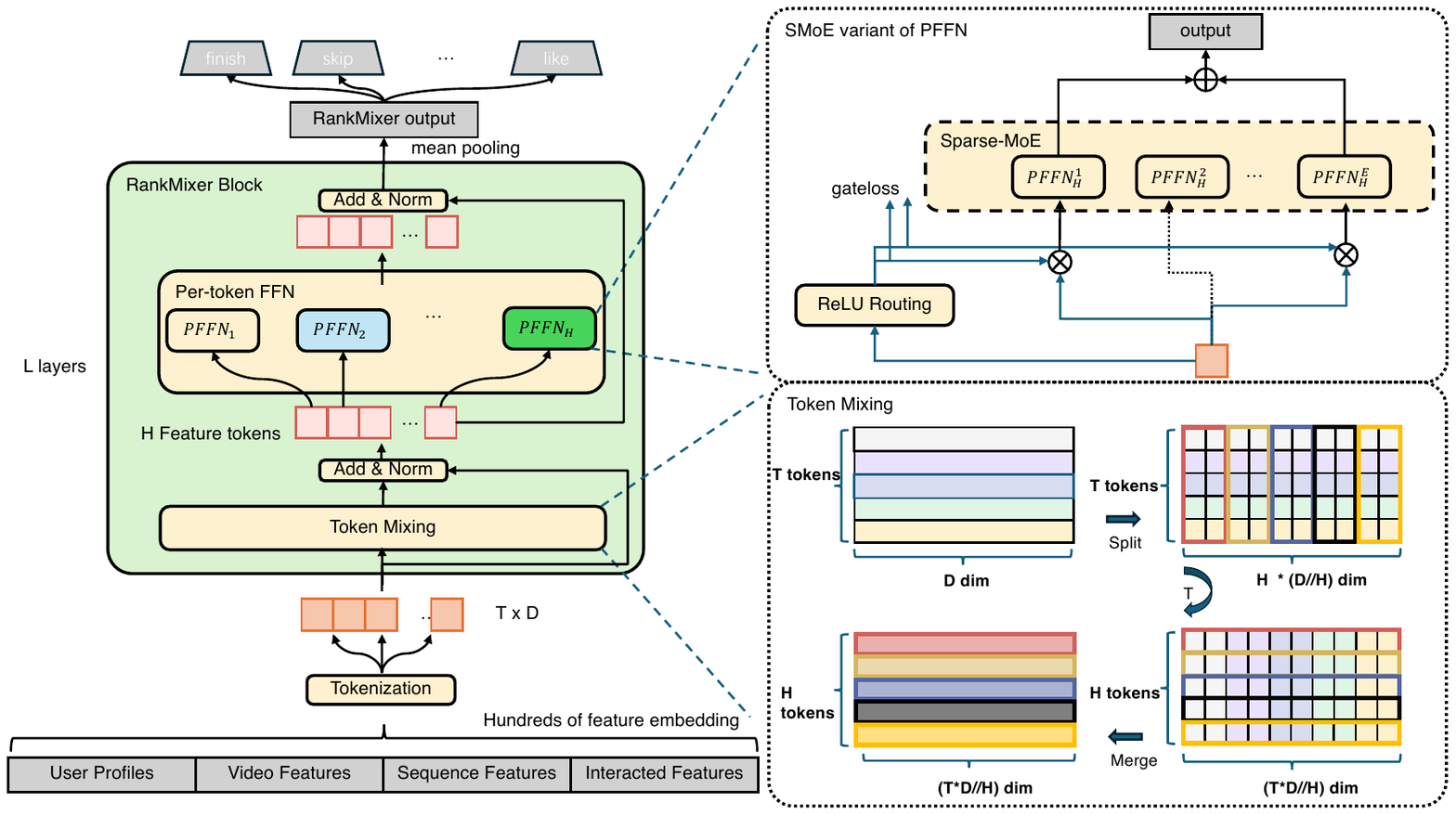}
    \caption{The architecture of a RankMixer block. A RankMixer block consists of two modules: \textbf{Multi-head Token Mixing} and \textbf{SMoE based Per-token FFN}. The token-mixing module divides each token’s embedding into $H$ smaller parts (heads), then recombines these parts across tokens to create new mixed tokens. This allows information from different features to interact with each other.}
    \label{fig:arch_overall}
\end{figure*}

% \begin{figure}
%     \centering
%     \includegraphics[width=0.8\linewidth]{img/RankMixer_overview_v2.pdf}
%     \caption{The overall architecture of RankMixer. First the input $\mathbf{e}_{\text{input}}$ is split into tokens. Each token encodes a group of feature embeddings with high semantic similarity. Then, $N$ RankMixer blocks are stacked to model feature interactions. The output of RankMixer $\mathbf{e}_{\text{output}}$ is the mean pooling of the output of the last RankMixer block.}
%     \label{fig:arch_overall}
% \end{figure}

 \subsection{Overall Architecture}
    RankMixer’s overall architecture consists of T input tokens processed by L successive RankMixer blocks, followed by an output pooling operator. Each RankMixer block has two main components: (1) a Multi-Head Token Mixing layer, and (2) a Per-Token Feed-Forward Network (PFFN) layer, as illustrated in Figure~\ref{fig:arch_overall} . 
    First, the input vector $\mathbf{e}_{\text{input}}$ is tokenized into $T$ feature tokens $\mathbf{x}_1, \mathbf{x}_2, ..., \mathbf{x}_T$, each representing a coherent feature vector.
    % we split the input $\mathbf{e}_{input}$ into $T$ tokens $\mathbf{x}_1, \mathbf{x}_2, ..., \mathbf{x}_T$ of the same dimension $D_{RankMixer}$. Each token encodes a group of feature embeddings with high semantic similarity. 
    RankMixer blocks iteratively refine token representations for $L$ layers through:
    % , the $n$-th RankMixer block is formulated as
    \begin{equation}
    \begin{split}
        \mathbf{S}_{n-1} &= {\rm LN}\left({\rm TokenMixing} \left( \mathbf{X}_{n-1} \right) +  \mathbf{X}_{n-1}\right),\\
         \mathbf{X}_{n} &= {\rm LN} \left({\rm PFFN} \left( \mathbf{S}_{n-1} \right)  +\mathbf{S}_{n-1} \right),
    \end{split}
    \end{equation}
    where ${\rm LN}(\cdot)$ is a layer normalization function, ${\rm TokenMixing}(\cdot)$ and ${\rm PFFN}(\cdot)$ are the multi-head Token Mixing module and the per-token FFN module, $\mathbf{X}_{n} \in \mathbb{R}^{T \times D}$ is the output of the $n$-th RankMixer block, $\mathbf{X}_{0} \in \mathbb{R}^{T \times D}$ is stacked by $\mathbf{x}_1, \mathbf{x}_2, ..., \mathbf{x}_T$, and $D$ is the hidden dimension of the model. 
    The output representation $\mathbf{o}_{\text{output}}$ derives from the mean pooling of final layers's representations $\mathbf{X}_{L}$ which will be used to compute the different task's predictions.  

\subsection{Input Layer and Feature Tokenization}

The first thing to build a large-scale recommendation model is to prepare inputs with abundant information, such as \textbf{User features} include user ID and other user information, \textbf{candidate features} include video ID, author ID etc., \textbf{Sequence Features}  processed through Sequence Module \cite{din,longer} to capture temporal interests, yielding $\mathbf{e}_s$ and \textbf{Cross Features}. the cross features between the user and the candidate. All the features will be transformed into embeddings with diverse dimensions.  

To enable efficient parallel computation in later stages, embeddings of varying dimensions must be transformed into dimension-aligned vectors, called feature-tokens. We refer to this embedding alignment process as tokenization. The simplest strategy is assigning one embedding per feature which introduces several challenges given typically hundreds of features. Having hundreds of tokens inevitably reduces the parameters and computation allocated per token to small fragments, resulting in insufficient modeling of important features and under-utilization of GPU cores. Conversely, too few tokens (e.g. a single token) degrade the model structure into a simple Deep Neural Network (DNN), unable to distinctly represent diverse feature spaces, which risks dominant features overshadowing others.

To overcome these issues, we propose a semantic-based tokenization approach with domain knowledge and group features into several semantically coherent clusters. These grouped features are sequentially concatenated into one embedding vector $e_{\text{input}} \;=\; [\,e_{1}; e_{2}; \dots; e_{N}\,]$, and subsequently partitioned into an appropriate number of tokens with fixed dimension sizes. Each feature token $x_i \in \mathbb{R}^{D}$ captures a set of feature embeddings that represent a similar semantic aspect. 
\begin{equation}
  \newline
  \text{x}_{i} \;=\; {\rm Proj}(e_{\text{input}}\bigl[d\cdot(i-1) : d\cdot i\bigr]),
  \quad i = 1,\dots,T,
  \label{eq:tokenization}
\end{equation}
where \(e_{\text{input}}\) is the concatenated embedding vector,
\(d\) is the fixed dimension per token, \(N\) is the number of
feature-groups, and \(T\) is the resulting token count and the ${\rm Proj}$ function maps the splitted embedding into $D$ dimension.

  \subsection{RankMixer Block}
    % We will introduce how to split $\mathbf{e}_{input}$ into tokens, the multi-head Token Mixing module and the per-token FFN module in the following parts.
    
  % \subsubsection{Tokenization}

  %   In large-scale industrial recommendation models, the input dimension $D_{\text{input}}$ is ofen very large, making direct application of deep MLPs to the input vector $\mathbf{e}_{\text{input}}$ computationally expensive and ineffective. 
  %   % Additionally, the model may struggle to learn meaningful feature interactions with such a large fully connected network.
  %   A common approach to address this problem is to split the input into tokens before applying feature interactions, such as DHEN~\cite{zhang2022dhen} and AutoInt~\cite{autoint}, where the number of tokens corresponds to the total number of features. However, this method can create critical limitations.
  %   Feature embeddings often exhibit varying dimensions due to their intrinsic information density, which poses two challenges. First, uniformly aligning embeddings to the maximum dimension results in excessive padding. Second, projecting all feature embeddings into the same dimension generates numerous small matrix operations, leading to under-utilization of GPU cores.

  \subsubsection{Multi-head Token Mixing}

    % After the inner-token feature interaction, we design the multi-head Token Mixing module to enable effective information exchange across tokens.
    To facilitate effective information exchange across tokens which is important for feature cross and global information modeling, we introduce the multi-head Token Mixing module.
    Each token is evenly divided into $H$ heads, with the $h$-th head of token $\mathbf{x}_t$ denoted as $x_t^h$:
    \begin{equation}
        \Bigl[\,
             \mathbf{x}_t^{(1)}
             \;\Vert\;
             \mathbf{x}_t^{(2)}
             \;\Vert\;\dots\;\Vert\;
             \mathbf{x}_t^{(H)}
            \Bigr] = {\rm SplitHead} \left( \mathbf{x}_t \right). 
    \end{equation}
    These heads can be seen as projections of the token $\mathbf{x}_t$ into a lower-dimensional feature subspace since recommendation is task taking different perspectives into considerations.
    % $x_t^h$ can be considered as a projection from the token $t$ into a subspace:
    Token Mixing is used to fuse these sub-space vector for global feature interactions. Formally, the $h$-th token $\mathbf{s}^{h}$ corresponding to the $h$-th head after the multi-head Token Mixing is constructed as:
    \begin{equation}
        \mathbf{s}^{h} = {\rm Concat}\left( \mathbf{x}_1^{h}, \mathbf{x}_2^{h}, ..., \mathbf{x}_T^{h} \right).
    \end{equation}
    The output of the multi-head Token Mixing module is $\mathbf{S} \in \mathbb{R}^{H \times \frac{TD}{H}}$, which is stacked by all the shuffled tokens $\mathbf{s}_{1}, \mathbf{s}_{2}, ..., \mathbf{s}_{H}$. In this work, we set $H=T$ to maintain the same number of tokens after Token Mixing for residual connection.

    % We summarize the multi-head Token Mixing operation as follows:
    
    % \begin{equation}
    %     {\rm TokenMixing} \left( \mathbf{x}_1, \mathbf{x}_2, ..., \mathbf{x}_T \right) = {\rm Merge}\left( {\rm SplitHead} \left( \mathbf{x}_1, \mathbf{x}_2, ..., \mathbf{x}_T \right) \right).
    % \end{equation}
    
    After a residual connection and normalization module we can generate
    \begin{equation}
        \mathbf{s}_1, \mathbf{s}_2, ..., \mathbf{s}_T = {\rm LN}\left({\rm TokenMixing} \left( \mathbf{x}_1, \mathbf{x}_2, ..., \mathbf{x}_T \right) + \left( \mathbf{x}_1, \mathbf{x}_2, ..., \mathbf{x}_T \right)\right),
    \end{equation}
    
    Although self-attention has proven highly effective in large language models, we find it to be suboptimal for recommendation systems.
    % In recommendation tasks, the feature spaces are inherently heterogeneous, with user demographics, item attributes, and interaction contexts existing in distinct semantic and dimensional domains.
    % This contrasts sharply with NLP paradigms, where all tokens 
    In self-attention, attention weights are calculated using the inner product of tokens. 
    This method works well for NLP, where all tokens share a unified embedding space. 
    However, in recommendation tasks, the feature spaces are inherently heterogeneous.
    Computing an inner-product similarity between two heterogeneous semantic spaces is notoriously difficult—particularly in recommender systems, where the ID space of the features from user and item side may contain hundreds of millions of elements.
    Consequently, applying self-attention to such diverse inputs does not outperform the parameter-free multi-head Token Mixing approach and consumes more computations, Memory IO operations and GPU memory usage.
    
  \subsubsection{Per-token FFN}
    Previous DLRM and DHEN models tend to mix features from many disparate semantic spaces in a single interaction module, which may causes high-frequency fields dominate drowning out lower-frequency or long-tail signals and ultimately hurting overall recommendation quality.
    We introduce a parameter-isolated feed-forward network architecture, termed per-token FFN. In traditional designs, the parameters of FFN are shared across all tokens, but our approach processes each token with dedicated transformations, thus isolating the parameters for each token. For the $t$-th token $\mathbf{s}_t$, the per-token FFN can be expressed as
    \begin{equation}
        \mathbf{v}_t = f_{\text{pffn}}^{t,2} \left({\rm Gelu} \left(f_{\text{pffn}}^{t,1} \left(\mathbf{s}_t \right) \right) \right),
    \end{equation}
    where
    \begin{equation}
        f_{\text{pffn}}^{t,i}(\mathbf{x})=\mathbf{x}\mathbf{W}_{\text{pffn}}^{t, i} + \mathbf{b}_{\text{pffn}}^{t, i}
    \end{equation}
    % where $f_{pffn}^{t,i}(\cdot)$ 
    is the $i$-th layer MLP of the per-token FFN, 
    % $f_{pffn}^{t,i}(\mathbf{x})=\mathbf{x}\mathbf{W}_{pffn}^{t, i} + \mathbf{b}_{pffn}^{t, i}$, 
    $\mathbf{W}_{\text{pffn}}^{t, 1} \in \mathbb{R}^{D \times kD}$, 
    $\mathbf{b}_{\text{pffn}}^{t, 1} \in \mathbb{R}^{kD}$,
    $\mathbf{W}_{\text{pffn}}^{t, 2} \in \mathbb{R}^{kD \times D}$, 
    $\mathbf{b}_{\text{pffn}}^{t, 2} \in \mathbb{R}^{D}$,
    $k$ is a hyperparameter to adjust the hidden dimension of the per-token FFN,
    ${\rm Gelu(\cdot)}$ is the Gelu activation function, $\mathbf{s}_t \in \mathbb{R}^{D}$ is the $t$-th token.

    We summarize the Per-token FFN module as
    \begin{equation}
        \mathbf{v}_1, \mathbf{v}_2, ..., \mathbf{v}_T = {\rm PFFN}\left( \mathbf{s}_1, \mathbf{s}_2, ..., \mathbf{s}_T \right), 
    \end{equation}
    where
    \begin{equation}
        {\rm PFFN}\left( \mathbf{s}_1, \mathbf{s}_2, ..., \mathbf{s}_T \right) = f_{\text{pffn}}^{t,2} \left({\rm Gelu} \left(f_{\text{pffn}}^{t,1} \left(\mathbf{s}_1, \mathbf{s}_2, ..., \mathbf{s}_T \right) \right) \right).
    \end{equation}
Compared to the parameter-all-shared FFN, per-token FFN enhances the modeling ability by introducing more parameters while keeping the computational complexity unchanged.

It is worth emphasizing that a Per-token FFN differs from MMoE experts in that each Per-token FFN sees a distinct token input, whereas all experts in a MMoE share the same input.
Unlike a MMoE, where many experts process the same input, and unlike a Transformer, where different input share one FFN, RankMixer splits the inputs and the parameters simultaneously which is good for learning diversity in different feature sub-spaces.

 \subsection{Sparse MoE in RankMixer}
To further increase the scaling ROI, we can replace the dense FFNs of each pertoken  with
\emph{Sparse Mixture-of-Experts} (MoE) blocks, so that the model's capacity grows while computation cost
remains roughly constant.
Vanilla Sparse-MoE, however, degrades in RankMixer because:
(i) \textbf{uniform $k$-expert routing}.  Top-$k$ selection treats all feature tokens
equally, wasting the budget on low-information tokens and starving high-information ones, which hinders the model from capturing differences between tokens.
(ii) \textbf{expert under-training}.  Per-token FFNs already multiply parameters by
\#tokens; adding non-shared experts further explodes the expert count, producing
highly unbalanced routing and poorly trained experts;

We combine two complementary training strategies to solve the above problems.

%--------------------------- 1. ReLU-routed Sparse-MoE --------------------------------
\paragraph{ReLU Routing}
To grant tokens flexible expert counts and maintain differentiability, we replace the common
\(\text{Top}k {+} {\rm softmax} \) with a ReLU gate plus an adaptive $\ell_1$ penalty
\cite{Wang2024remoe}. Given the $j$-th expert $e_{i,j}(\cdot)$ for token $s_i\!\in\!\mathbb{R}^{d_h}$ and router $h(\cdot)$:
\begin{equation}
  G_{i, j} = {\rm ReLU} \bigl(h(\mathbf{s}_i)\bigr), \quad
  \mathbf{v}_i = \sum_{j=1}^{N_e} G_{i, j}\, e_{i, j}(\mathbf{s}_i),
  \label{eq:dsmoe_forward}
\end{equation}
where $N_e$ is the number of experts per token, $N_t$ is the number of tokens.
% \begin{equation}
%   R(x)= {\rm ReLU} (sW)\in\mathbb{R}^{E},
%   \label{eq:relu_gate}
% \end{equation}
% where $W\!\in\!\mathbb{R}^{d_h\times E}$.  
ReLU Routing will activate more experts for high-information tokens and improve the parameter efficiency.
Sparsity is steered by
\(\mathcal{L}_{\mathrm{reg}}\) with an coefficient~$\lambda$ that
keeps the average active-expert ratio near the budget:
\begin{equation}
  \mathcal{L}= \mathcal{L}_{\text{task}}
              +\lambda\,\mathcal{L}_{\mathrm{reg}},
  \quad
  \mathcal{L}_{\mathrm{reg}}
  =\sum_{i=1}^{N_t}\sum_{j=1}^{N_e} G_{i,j}.
  \label{eq:relu_loss}
\end{equation}

%-------------------------- 2. Dense-training / Sparse-inference -----------------------
\paragraph{Dense-training / Sparse-inference (DTSI-MoE)}

Inspired by \cite{pan2024ads}, two routers $h_{\text{train}}$ and $h_{\text{infer}}$ are adopted, and $\mathcal{L}_{\mathrm{reg}}$ is applied only to $h_{\text{infer}}$. Both $h_{\text{train}}$ and $h_{\text{infer}}$ are updated during training, while only $h_{\text{infer}}$ is used in inference. It turns out that DS-MoE makes experts do not suffer from under-training while reducing inference cost.

% \begin{equation}
%   \mathbf{v}_\text{sparse\_inference}
%   =\sum_{e:R(x)_e>0}R(x)_e\,e_e(\mathbf{s}).
%   \label{eq:relu_inf}
% \end{equation}

% For a token $s\!\in\!\mathbb{R}^{d_h}$, training router $h_{train}(\cdot)$, and experts $e_i(\cdot)$:
% \begin{equation}
%   G_i = {\rm ReLU} \bigl(h_{train}(s)\bigr), \quad
%   \mathbf{v}_\text{train} = \sum_{i=1}^{N} G_i\,e_i(\mathbf{s}).
%   \label{eq:dsmoe_forward}
% \end{equation}
% In inference we keep only a few experts using the following Dynamic Routing strategies.
% % \begin{equation}
% %   \mathbf{v}=\sum_{i\in\TopK(x)} G_i\,e_i(\mathbf{s}).
% %   \label{eq:sparse_inf}
% % \end{equation} 

 \subsection{Scaling Up Directions}
 RankMixer is intrinsically a highly parallel and extensible
architecture.  Its parameter count and computational cost can be
expanded along four orthogonal axes: Token count \textbf{\(T\)}, Model width \textbf{\(D\)}, Layers \textbf{\(L\)} and Expert Numbers \textbf{\(E\)}. 
For full-dense-activated version, parameter and forward FLOPs for one sample can be computed as
\begin{equation}
  \text{\#Param}_{}
  \;\approx\;
  2\,k\,L\,T\,D^{2},
  \qquad
  \text{FLOPs}_{}
  \;\approx\;
  4\,k\,L\,T\,D^{2},
  \label{eq:dense_cost}
\end{equation}
The $k$ is the scale ratio adjusting the hidden dimension of FFN.
In the \emph{Sparse-MoE} version, the \emph{effective} parameters and
compute per token are further scaled by the sparsity ratio~$s=\frac{\text{\#Activated\_Param}}{  \text{\#Total\_Param}}$.
\section{Experiments}
\label{sec:experiments}

% In this section, we explore the following key research questions:

% \begin{itemize}[leftmargin=10pt]
%     \item \textbf{RQ1:} : How does RankMixer perform compared with state-of-the-art models in the comparable scale and complexity.
%     \item \textbf{RQ2:} What is the performance scaling curves  when scaling different model structures and which model's scaling law is the most steepest
%     \item \textbf{RQ3:} How does the key commponet in RankMixer affecting its scaling performance
%     \item \textbf{RQ4:} How to analyze and optmize RankMixer model's cost and efficiency in the real-world application.
% \end{itemize}
\subsection{Experiment Settings}

\subsubsection{Datasets and Environment}

The offline experiments were conducted using the training data from the Douyin recommendation system. These data are derived from Douyin's online logs and user feedback labels. The training dataset includes over 300 features, such as numerical features, ID features, cross features, and sequential features, involving billions of user IDs and hundreds of millions of video IDs, which are all converted into embeddings. The data covers trillions of records per day, and experiments were conducted on data collected over a two-week period.

\subsubsection{Evaluation Metrics}

To evaluate model performance, we use AUC (Area Under the Curve) and UAUC (User-level AUC) as the primary performance metrics and Parameter count, FLOPs and MFU as the efficiency metrics listed as follows: \textbf{Finish/Skip AUC/UAUC:} A finish=1/0 or skip=1/0 label indicates whether a user completed watching a video or slide to the next video in a short-time. We evaluate the AUC and UAUC for this finish label. an AUC increase of 0.0001 can be regarded as a confidently significant improvement. \textbf{Dense-Param:} The number of parameters in the dense part, excluding the sparse embedding parameters.  Training \textbf{Flops/Batch:} The number of floating-point operations (FLOPs) required to run a single batch of 512 through the model, representing the computational cost of training.  \textbf{MFU:} MFU(Model FLOPs Utilization) is a metric that measures how effectively a model utilizes floating-point operations provided by the hardware, calculated by dividing the model’s actual FLOPs consumption by the hardware’s theoretical FLOPs capacity.

\subsubsection{Baselines}

We compare against the following widely recognized SOTA baselines: \textbf{DLRM-MLP}:which is the vanilla MLP for feature crossing as the experiment baseline, \textbf{DCNv2}\cite{wang2021dcn},\textbf{RDCN}\cite{DBLP:conf/kdd/BorisyukZSZTPDH24} the Sota of feature cross model. \textbf{MoE} scales up by using mutliply parallel experts. \textbf{AutoInt}\cite{song2019autoint}, \textbf{Hiformer}\cite{gui2023hiformer} combines heterogeneous self-attention layer and low-rank approximation matrix computation. \textbf{DHEN}\cite{zhang2022dhen}: which combines different kind of feature-cross block and stacks multiple layers (DCN/self-attention/FM/LR) \textbf{Wukong}\cite{zhangwukong} investigates the scaling law of feature interaction following the arch of DHEN with Factorization Machine Block (FMB) and Linear Compress Block (LCB).

All experiments were conducted on hundreds of GPUs in a hybrid distributed training framework that the sparse part is updated asynchronously, while the dense part is updated synchronously. The optimizer hyperparameters were kept consistent across all models. For the dense part, we used the RMSProp optimizer with a learning rate of 0.01, while the sparse part used the Adagrad optimizer. 

\subsection{Comparison with SOTA methods}
% 需要在导言区引入:
% \usepackage{graphicx}
% \usepackage{booktabs}

\begin{table}[t]
\centering
\caption{Performance \& efficiency comparison of $\sim$100 M-parameter recommendation models (best values in \textbf{bold}).}
\label{table:main_result}
\resizebox{\linewidth}{!}{%
\renewcommand\arraystretch{1.05}
\begin{tabular}{lcccccc}
\toprule
\multirow{2}{*}{Model} & \multicolumn{2}{c}{\textbf{Finish}} & \multicolumn{2}{c}{\textbf{Skip}} & \multicolumn{2}{c}{\textbf{Efficiency}} \\
\cmidrule(lr){2-3} \cmidrule(lr){4-5} \cmidrule(l){6-7}
 & AUC$\uparrow$ & UAUC$\uparrow$ & AUC$\uparrow$ & UAUC$\uparrow$ & Params & FLOPs/Batch \\
\midrule
DLRM-MLP (base)     & 0.8554 & 0.8270 & 0.8124 & 0.7294 & 8.7 M  & 52 G \\
\midrule
DLRM-MLP-100M       & +0.15\% & —        & +0.15\% & —        & 95 M  & 185 G \\
DCNv2               & +0.13\% & +0.13\% & +0.15\% & +0.26\% & 22 M  & 170 G \\
RDCN                & +0.09\% & +0.12\% & +0.10\% & +0.22\% & 22.6 M & 172 G \\
MoE                 & +0.09\% & +0.12\% & +0.08\% & +0.21\% & 47.6 M & 158 G \\
AutoInt             & +0.10\% & +0.14\% & +0.12\% & +0.23\% & 19.2 M & 307 G \\
DHEN                & +0.18\% & +0.26\% & +0.36\% & +0.52\% & 22 M  & 158 G \\
HiFormer            & +0.48\% & —        & —        & —        & 116 M & 326 G \\
Wukong              & +0.29\% & +0.29\% & +0.49\% & +0.65\% & 122 M & 442 G \\
\midrule
\textbf{RankMixer-100M} & \textbf{+0.64\%} & \textbf{+0.72\%} & \textbf{+0.86\%} & \textbf{+1.33\%} & 107 M & 233 G \\
\textbf{RankMixer-1B}   & \textbf{+0.95\%} & \textbf{+1.22\%} & \textbf{+1.25\%} & \textbf{+1.82\%} & \textbf{1.1 B} & \textbf{2.1T} \\
\bottomrule
\end{tabular}%
}
\end{table}

To explore how to scale up the model, we compare models with similar parameter sizes around 100 million to determine which model structure performs best with the same computational cost.

The main results of the performance of our method and baselines
are summarized in  Table~\ref{table:main_result}.
 We can observe RankMixer significantly outperforms other SOTA models across multiple objectives and metrics.

We take a closer look at each model. First, simply scaling DLRM up to 100 million parameters yields only limited gains, underscoring the necessity of designing models tailored to recommendation data characteristics for better scaling performance. We then compare the RankMixer model with other classic cross-structure designs such as DCN, RDCN, AutoInt, and DHEN, and find they suffer from an imbalance between model parameters and computational cost. Even with relatively small parameter sizes, these models already exhibit large FLOPs, indicating design shortcomings that constrain their results in the table. Meanwhile, although RankMixer achieves the best performance, its FLOPs remain relatively moderate among all the models when scaled to 100 million parameters, reflecting a balanced approach to model capacity and computational load. 

We also compare RankMixer with several commonly used state-of-the-art scaling-up models—like Hiformer and wukong—and find that under similar parameter settings, RankMixer not only performs better but also has lower computational requirements.

\subsection{Scaling Laws of different models}

\begin{figure}
    \centering
    \includegraphics[width=\linewidth]{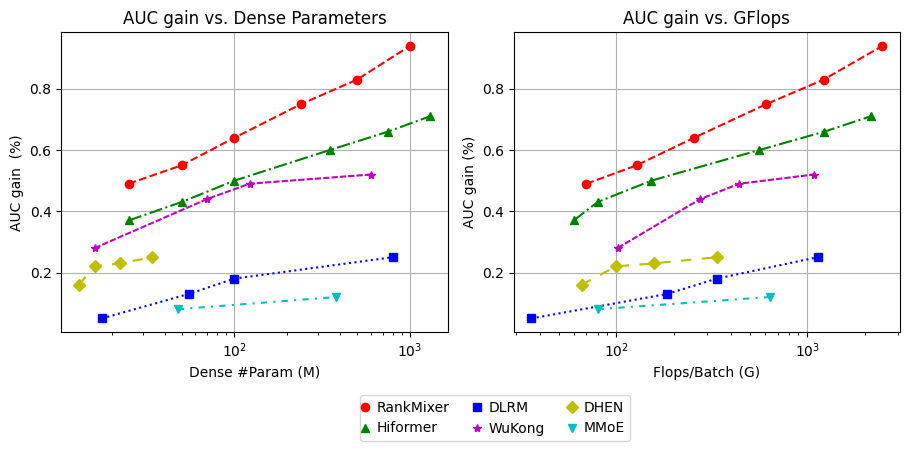}
    \caption{Scaling laws between finish Auc-gain and Params/Flops of different models.  The x-axis uses a logarithmic scale.}
    \label{fig:scaling_law}
\end{figure}

In the Figure \ref{fig:scaling_law}, we show the scaling law curves observed from both parameter size and Flops. The RankMixer model shows the steepest scaling law both interms of parameters, and FLOPs. RankMixer is consistently superior to the other models. 

Although the wukong model exhibits a relatively steep parameter curve, its computational cost increases even more rapidly; as a result, the gap between it and both RankMixer and hiformer is even larger on the AUC vs FLOPs curve. Moreover, hiformer’s performance is slightly inferior to RankMixer’s, reflecting that its reliance on token segmentation at a feature level and on Attention affects its efficiency. The scaling of DHEN is not ideal, reflecting the limited scalability of its cross structure. Moreover, MoE’s strategy of scaling by adding experts brings about challenges in maintaining expert balance, which results in suboptimal scaling performance.

To be Sepcific, We can scale up RankMixer model by increasing width ($D$), feature token ($T$) and Layer($L$). In our experiments we observed a conclusion shared with LLM scaling laws: model quality correlates primarily with the total number of parameters, and different scaling directions (depth L, width D, tokens T) yield almost identical performance. From a compute-efficiency standpoint, while larger hidden-dim generate larger matrix-multiply shapes and thus achieve higher MFU than stacking more layers. So the final configuration for 100M and 1B are set as ($D$=768, $T$=16, $L$=2) and ($D$=1536, $T$=32, $L$=2) respectively.

\subsection{Ablation Study}

\begin{table}[h]
\centering
\caption{Ablation on components of RankMixer-100M}
\label{ablation study}
\small
\begin{tabular}{@{\quad}l c@{\quad}}
\toprule
\textbf{Setting} & $\Delta$AUC \\ 
\midrule
w/o skip connections                          & $-0.07\%$ \\
w/o multi-head token mixing                   & $-0.50\%$ \\
w/o layer normalization                       & $-0.05\%$ \\
Per-token FFN $\rightarrow$ shared FFN        & $-0.31\%$ \\
\bottomrule
\end{tabular}
\end{table}

%-------------------------------------------------------
% Table 2 – routing–strategy comparison
\begin{table}[h]
\centering
\caption{Token2FFN Routing–strategy comparison}
\label{ablation-routing}
\small
\begin{tabular}{@{\quad}l c c c@{\quad}}
\toprule
\textbf{Routing strategy} & $\Delta$AUC & $\Delta$Params & $\Delta$FLOPs \\ 
\midrule
All-Concat-MLP & $-0.18\%$ & 0.0\% & 0.0\% \\
All-Share      & $-0.25\%$ & 0.0\% & 0.0\% \\
Self-Attention & $-0.03\%$ & +16\% & +71.8\% \\
\bottomrule
\end{tabular}
\end{table}

In the RankMixer-100M model, we performed ablation studies on residual connections, Multi-Head Token-Mixing. From Table \ref{ablation study}, we can see that removing these components significantly decreased the model's performance. Removing Multi-Head Token-Mixing loses global information, as each FFN only models partial features without interaction. Removing residual connections and LayerNorm also worsens performance, reducing training stability and making gradient explosion or vanishing issues more likely.
% \include{table/online_abtest}
% \begin{table}[h]
%     \centering
%     \caption{Token Routing Strategies}
%     \begin{tabular}{|c|c|c|}
%         \hline
%         Strategy & AUC & Flops \\
%         \hline
%         All-Concat-MLP & 0.XX & X \\
%         Self-Attention & 0.XX & X \\
%         All-Share & 0.XX & X \\
%         Multi-Head Token-Mixing & 0.XX & X \\
%         \hline
%     \end{tabular}
% \end{table}
        
We further analyzed the token mixing strategies, i.e., the routing strategies from feature tokens to FFNs in the Table \ref{ablation-routing}. The routing strategies compared with Multi-Head Token Mixing (Multi-Head Token-Mixing) include: \textit{All-Concat-MLP:} Concatenates all tokens and processes them through a large MLP before splitting them into the same number of tokens. The decrease of performance shows the challenges in learning large matrices and weakening local information learning. \textit{All-Share:} No splitting, the entire input vector is shared and fed to each per-token FFN similar as MoE. The performane declines siginifantly which show the importance of feature subspace split and independent modeling contrast to all-shared input.
\textit{Self-Attention:} Applies self-attention between tokens for routing. Its performance is slightly inferior to the Multi-Head Token-Mixing and also suffers from high computational cost, which shows the difficulty of learning similarity across hundreds of different feature subspaces. 

\subsection{Sparse-MoE Scalability and Expert Balance}
\label{sec:smr_results}

\begin{figure}
    \centering
    \includegraphics[width=\linewidth]{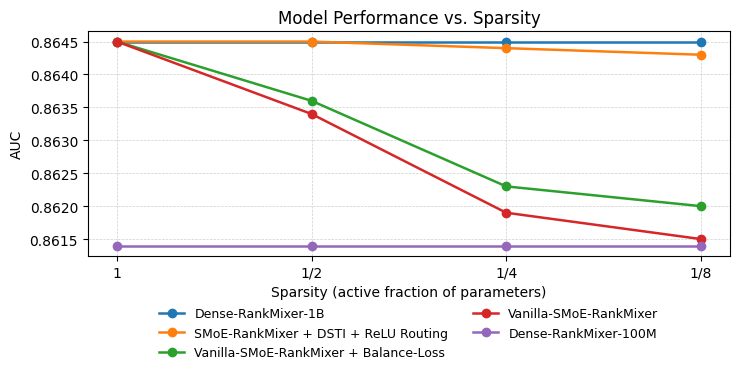}
    \caption{AUC performance of RankMixer variants under decreasingly sparse activation ratio (1,1/2,1/4,1/8 of experts): dense-training + ReLU-routed SMoE preserves almost all accuracy of the 1 B dense model.}
    \label{fig:sparsity_curve}
\end{figure}

\paragraph{Scalability.}
Figure~\ref{fig:sparsity_curve} plots offline \textsc{AUC} gains versus Sparsity of SMoE. Combining Dense-Training-Sparse-Inference with ReLU routing is essential for preserving accuracy under aggressive sparsity, enabling RankMixer to scale parameter capacity (and memory footprint) by > 8× with nearly no loss in AUC and with substantial inference-time savings (+50\% improvement over throughput).  Vanilla SMoE's performance drops monotonically as fewer experts are activated, illustrating the expert-imbalance and under-training issues we identified. Adding a load-balancing loss reduces the degradation relative to vanilla SMoE, yet still falls short of the DTSI + ReLU version because the problems mostly lies in the expert training instead of the router.
This validates Sparse-MoE as the
path to scale RankMixer from the current 1\,B parameters to future
10\,B-scale deployments without breaching the cost budgets.

\begin{figure}
    \centering
    \includegraphics[width=0.7\linewidth]{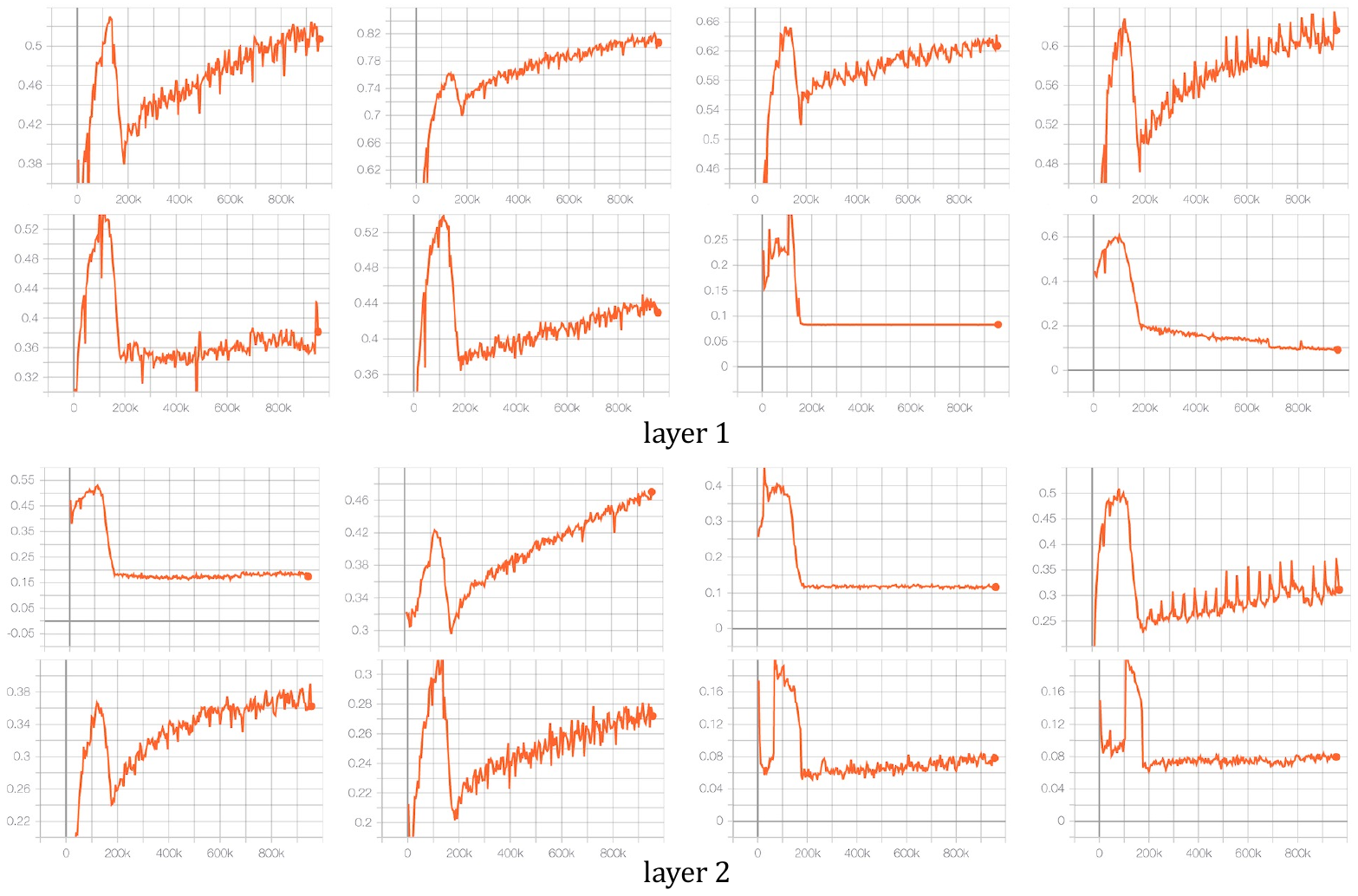}
    \caption{activated expert ratio for different {\text{token}} in {\text{RankMixer}}.}
    \label{fig:expert_balance}
\end{figure}

\begin{table*}[h]
\caption{Online AB test result for Feed Recommendation Scenarios in both Douyin and Douyin lite app and the improvements are all statistically significant. According to long-term reverse A/B tests and continuous observations over long-term reverse experiments, the gains provided by RankMixer-1B have not yet converged, and the results presented here are still improving.}
\resizebox{\textwidth}{!}{
\renewcommand\arraystretch{1.08}
\begin{tabular}{lcccclccccl}
\hline
                       & \multicolumn{5}{c}{\textbf{Douyin app}}                                                      & \multicolumn{4}{c}{\textbf{Douyin lite}}                                  &                  \\ \cmidrule(lr){2-6}\cmidrule{7-11} 
                       & \textbf{Active Day}$\uparrow$ & \textbf{Duration}$\uparrow$ & \textbf{Like}$\uparrow$ & \textbf{Finish}$\uparrow$ & \textbf{Comment}$\uparrow$ & \textbf{Active Day}$\uparrow$ & \textbf{Duration}$\uparrow$ & \textbf{Like}$\uparrow$ & \textbf{Finish}$\uparrow$ & \textbf{Comment}$\uparrow$ \\ \hline

\textbf{Overall}       & +0.2908\%           & +1.0836\%         & +2.3852\%     & +1.9874\%       & +0.7886\%        & +0.1968\%           & +0.9869\%          & +1.1318\%     & +2.0744\%        & +1.1338\%        \\ \hline 
\textbf{Low-active}    & +1.7412\%            & +3.6434\%         & 
+8.1641\%     & +4.5393\%       & +2.9368\%         & +0.5389\%           & +2.1831\%         & +4.4486\%      & +3.3958\%       & +0.9595\%        \\
\textbf{Middle-active} & +0.7081\%           & +1.5269\%         & +2.5823\%     & +2.5062\%       & +1.2266\%         & +0.4235\%           & +1.9011\%         & +1.7491\%      & +2.6568\%         & +0.6782\%        \\
\textbf{High-active}   & +0.1445\%           & +0.6259\%         & +1.828\%      & +1.4939\%       & +0.4151\%          & +0.0929\%           & +1.1356\%         & +1.8212\%      & +1.7683\%        & +2.3683\% 
\\ \hline
\end{tabular}
}
\label{online_ab}
\end{table*}

\begin{table}[t]
\centering
\caption{Online lift of RankMixer in Advertising}
\label{tab:ads-search}
% \resizebox{0.9\linewidth}{!}{%
\footnotesize
\renewcommand\arraystretch{0.95}
\begin{tabular}{@{}lccc@{}}
\toprule
 & \multicolumn{2}{c}{\textbf{Advertising}}\\
\cmidrule(lr){2-3}
Metric & $\Delta$AUC$\uparrow$ & ADVV$\uparrow$  \\
\midrule
Lift  & +0.73\% & +3.90\%\\
\bottomrule
\end{tabular}%
\label{ad_search}
\end{table}

\begin{table}[h]
    \caption{Metrics of Online Model Deployment and Cost}
    \resizebox{0.48\textwidth}{!}{%
    \begin{tabular}{c|c|c|c}
        \hline
        Metric & OnlineBase-16M & RankMixer-1B & Change \\
        \hline
        \#Param & 15.8M & 1.1B & $\uparrow$ \textbf{70$\times$} \\
        FLOPs & 107G & 2106G & $\uparrow$ 20.7$\times$ \\
        Flops/Param(G/M) & 6.8 & 1.9 & $\downarrow$ 3.6$\times$ \\
        MFU & 4.47\% & \textbf{44.57}\% & $\uparrow$ 10$\times$ \\
        Hardware FLOPs & fp32 & fp16 & $\uparrow$ 2$\times$ \\
        Latency & 14.5ms & \textbf{14.3ms} & - \\
        \hline
    \end{tabular}%
    }
    \label{case_study}
\end{table}

\paragraph{Expert balance and diversity.}
Vanilla Sparse MoE often suffers from expert imbalance, which in turn leaves some experts under-trained and eventually leads to “dying experts” (experts that are almost never activated) and only a few fixed experts are constantly activated.
Figure~\ref{fig:expert_balance} shows that combining DTSI (dense-training, sparse-inference) with ReLU routing effectively resolves this issue: Dense-training guarantees that most expert receives sufficient gradient updates, preventing expert starvation.
ReLU routing makes the activation ratio dynamic across tokens—the activation proportion shows in the figure varies adaptively according to its information content, which aligns well with the diverse and highly dynamic distribution of recommendation data.
\subsection{Online Serving cost}

How can we prevent inference latency from exploding with a two-order-of-magnitude increase in parameters? In practical systems, latency is inversely proportional to throughput and directly proportional to the cost of serving machine resources. 
Compared to our previous fully-deployed 16M-parameter model (with a structure integrated with DLRM and DCN), our RankMixer model scaled parameters by approximately 70$\times$ to 1B. Despite this significant parameter increase, the final inference latency remained stable due to our hardware-aligned model design and optimization strategies.

When greatly increasing model parameters, latency can be decomposed into the following formula:
\[
\text{Latency} = \frac{\#\text{Param} \times \text{FLOPs/Param ratio}}{\text{MFU} \times \text{(Theoretical Hardware FLOPs)}}
\]
As illustrated in Table~6, the two-order-of-magnitude parameter increase is progressively counteracted by \textbf{3.6$\times$} decrease of FLOPs/Param ratio, \textbf{10\,$\times$} increase of MFU and Quatization-based \textbf{2\,$\times$} Hardware FlOPs improvements.

% 在线上实际部署对比我们在线全量的一个16M的模型（一个融合了DLRM和DCN 结构），我们1B的模型在参数量上提升了约70倍到1B，但通过hardware-aligned model design和优化，最终latency持平。

% 当大幅提升模型参数后，latency可以breakdown成以下公式
% latency = \frac{#Param * FLOPs/Param ratio} {MFU * (机器理论算力)}。
% 从表格6中可以看到，两个数量级的参数increase是如何被逐步拆解抵消的。
\begin{enumerate}[leftmargin=1.3em]
    \item \textbf{FLOPs/Param ratio.}  
          The three row of Table~\ref{case_study} reports the number of floating-point operations (FLOPs) required \emph{per parameter}.  
          Thanks to the model design, RankMixer achieves a 70-fold increase in parameters with only around a 20-fold increase in FLOPs, achieving only one-third the FLOPs/Param ratio of the baseline—a \textbf{3.6,$\times$ efficiency gain}. In other words, RankMixer can accommodate three times more parameters than the baseline at the same FLOPs budget.
    \item \textbf{Model FLOPs Utilization (MFU).}  
          Also shown in Table~\ref{case_study}, MFU indicates the utilization of machine computing.  
          By using large GEMM shapes, good parallel topology (fusing parallel per-token FFNs into one kernel), reduce the memory bandwith cost and overhead, RankMixer raises MFU by nearly \textbf{10\,$\times$}, shifting the model from an \textit{Memory-bound} to a \textit{Compute-bound} regime.
    \item \textbf{Quantization}
         Half-precision (fp16) inference will increase the theoretical peak Hardware FLOPS of GPUs by \textbf{2\,$\times$}.The primary computations in RankMixer consist of several large matrix multiplications that are well suited for half-precision (fp16) inference as mentioned before
\end{enumerate}

\subsection{Online Performance }
To verify the universality of RankMixer as a scaling
recommendation model framework, we conducted online experiments in the two
core application scenarios of personalised ranking—\emph{feed
recommendation} and \emph{advertising}, covering
major use-cases of personalized ranking.
For each scenario we monitor the following key performance indicators:
\begin{itemize}
  \item \textbf{Feed–Recommendation.}  
        \emph{Active Days} is he average number of active days per user during the experiment period, a substitute for DAU growth.;  
        \emph{Duration} measures cumulative stay time on the App;  
        \emph{Finish/Like/Comment:} User's total complete plays, likes, and comments.
  \item \textbf{Advertising.}  
        We report the model quality metrics ($\Delta\emph{AUC}$) and revenue metrics
         \emph{ADVV} (Advertiser Value).
  % \item \textbf{Search.}  
  %       Quality is gauged by query-level $\Delta\emph{AUC}$,
  %       \emph{Active Days}, and \emph{Query-change rate} which measures how often users modify their initial search queries during a search session and lower is better;.
\end{itemize}

% \begin{itemize}[leftmargin=10pt]
%     \item \textbf{Active Days:} The average number of active days per user during the experiment period, a substitute for DAU growth.
%     \item \textbf{Duration:} The total time user spend on the app.
%     \item \textbf{Finish/Like/Comment:} User's number of complete plays, likes, and comments.
% \end{itemize}

% \begin{figure}
%     \centering
%     \includegraphics[draft,width=0.8\linewidth]{img/wider_input.png}
%     \caption{Online A/B test: the AAD improvement across days}
%     \label{fig:scale_dim}
% \end{figure}
The Previous baselines in these scenarios are 16M parameter model combined with DLRM and DCN. We replaced the Dense part with the RankMixer-1B model improving a 0.7\% AUC. We conduct online A/B test experiments across Feed Recommendation and Advertising. The long-term observation of A/B test result on Feed Recommendation for 8 months  are shown on the Table \ref{online_ab} \footnote{This result was updated on July 24, 2025, since the long-term experiments indicating the gains had not yet saturated.}
\unskip and the result of Advertising is shown on the Table \ref{ad_search}.

% The online results show significant improvements in Active Days, Duration, and finish, like rate across multiple activeness user groups.

RankMixer was deployed and evaluated in the three personalised-ranking application including the Feed Recommendation (RankMixer-1B) and Advertising (RankMixer-1B). RankMixer delivered statistically significant uplifts on all business-critical metrics. We can also observe that the improvements on low-active users are largest on the Table \ref{online_ab} compared to other activeness level user groups with over 1.7412\% Active Days improvements, which demonstrates the model’s strong generalization capability.
These results demonstrate that the RankMixer as a unified backbone generalises reliably to different application scenarios.

\section{Conclusion}
\label{sec:conclusion}
In this article, we introduce our latest RankMixer model that has been fully deployed on Douyin Feed ranking. It combines the model designs for heterogeneous feature
interactions and the highly parallelizable architecture for serving efficiency. Experiments have demonstrated its superior performance and steep scaling law. After its full deployment on the Douyin app, it has obtained 0.3\% and  1\% increase on active days and App duartion.

\bibliographystyle{ACM-Reference-Format}
\bibliography{software}

\end{document}